\documentstyle[preprint,prl,aps,epsfig]{revtex}

\begin{document}

\title{Coherent Vector Meson Photoproduction with Nuclear Breakup in 
Relativistic Heavy Ion Collisions} 

\author{Anthony J. Baltz$^1$, Spencer R. Klein$^2$ and Joakim Nystrand$^3$} 

\address{$^1$Brookhaven National Laboratory, Upton, NY, 11973, USA\break
$^2$Lawrence Berkeley National Laboratory, Berkeley, CA 94720, USA\break
$^3$Dept. of Physics, Lund University, Lund SE-22100, Sweden} 

\break
\maketitle

\begin{abstract}
\vskip -.2 in 

Relativistic heavy ions are copious sources of virtual photons.  The
large photon flux gives rise to a substantial photonuclear interaction
probability at impact parameters where no hadronic interactions can
occur. Multiple photonuclear interactions in a single collision are
possible. In this letter, we use mutual Coulomb excitation of both
nuclei as a tag for moderate impact parameter collisions.  We
calculate the cross section for coherent vector meson production
accompanied by mutual excitation, and show that the median impact
parameter is much smaller than for untagged production.  The vector
meson rapidity and transverse momentum distribution are very different
from untagged exclusive vector meson production.

\end{abstract}

\narrowtext


The large charge ($Z$) of heavy nuclei gives rise to strong electromagnetic 
fields. For relativistic nuclei, these fields may be treated as an 
almost-real virtual photon beam, following the Weizs\"acker-Williams 
method. In relativistic heavy ion collisions, the photon field of one 
nucleus can produce a photonuclear interaction in the other. Many types
of photonuclear interactions are possible: nuclear excitation,
coherent or incoherent vector meson production, and other incoherent
photonuclear interactions\cite{reviews}. When the impact parameter $b$
is significantly larger than twice the nuclear radius $R_A$, hadronic
interactions are not possible, and photonuclear reactions can be cleanly
detected.  When $b > ~2R_A$, the
absolute probability for many of the photonuclear reactions can be 
substantial, and consequently, multiple reactions are possible. For
example, each ion can emit a photon, exciting the other nucleus, 
leading to mutual excitation\cite{mutualdissoc}\cite{BCW}\cite{mut2}.  
The excited ions typically decay by emission of one or more neutrons   
which move in the longitudinal direction with approximately the same 
momentum as the beam. This process has a distinctive event signature.

In this Letter, we consider vector meson production accompanied by 
mutual Coulomb dissociation, as shown in Figure 1.  More generally, 
we discuss the use of mutual dissociation as a `tag' of impact 
parameter. The vector meson calculations follow the model for 
exclusive production in Ref. \cite{usrates}. We will show that vector 
meson production accompanied by mutual excitation has a different 
impact parameter distribution than exclusive production, and this 
alters both the rapidity and transverse momentum distributions. 

\begin{figure}
\setlength{\epsfysize=0.3\textheight}
\center{\epsffile{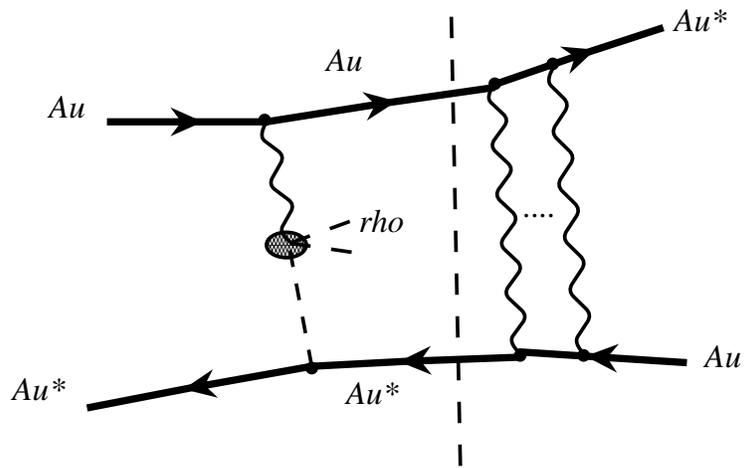}}
\caption[]{The dominant Feynman diagrams for 
vector meson production with nuclear excitation.}
\label{feynman}
\end{figure}

The diagram in Figure \ref{feynman}, corresponding to exchange of 3 photons, 
is dominant for vector meson production in coincidence with Coulomb breakup. 
For factorization to hold, these photons must be emitted independently. This 
was demonstrated by  S. N. Gupta\cite{gupta}. 

Calculations in Ref. \cite{hencken} have shown that the probability of
a photon leaving the emitting nucleus in an excited
state is small for the case of heavy ion two-photon particle production.
It is even less likely that the scattering represeneted by the dashed line in
Fig. \ref{feynman} will leave the emitting nucleus in an excited state.
One important difference between two-photon interactions and coherent
vector meson production is that the meson scatters from neutrons as
well as protons. In a GDR the protons and
neutrons oscillate against each other, so a force with equal coupling 
to protons and neutrons  
should not excite a GDR. The equi-coupling to neutrons and protons also 
holds for the $C=+$ mesons (mostly the $f_1$ and $a_2$) which can interact 
with a photon to produce vector mesons. The elastic scattering should thus 
not collectively excite the nucleus, and Fig. \ref{feynman} will be   
the dominant process.

These theoretical arguments are supported by data. The STAR
collaboration has observed $\rho$ production with and without nuclear
excitation, in gold-gold collisions at an energy of $\sqrt{s_{NN}}
= 130$ GeV per nucleon.  The $\rho$ data with excitation was collected
with a trigger based on mutual excitation, while the exclusive
$\rho$ sample was collected with a low-multiplicity charged particle 
trigger\cite{parkcity}. Both $\rho$ spectra are similarly peaked for 
$p_T <$~100~MeV/c.

The Weizs\"acker-Williams photon spectrum at a perpendicular distance $b$ from
the center of the emitting nucleus is~\cite{CahnJackson,Bauer,vidovic}:
\begin{equation}
{d^3n(b,k) \over dkd^2b}\! =\! {Z^2\alpha\over\pi^2 k}
\bigg|\!\int_0^\infty\!dk_\perp k_\perp^2 
{F^2(k^2_\perp + k^2/\gamma^2)k_\perp^2
\over \pi^2 (k_\perp^2 + k^2/\gamma^2)^2 }
J_1(bk_\perp)\bigg|^2
\label{photonflux}
\end{equation}
where $\alpha \sim 1/137$ is the electromagnetic coupling constant, $k$
is the photon energy, $k_\perp$ the photon transverse momentum with
respect to the direction of nuclear motion.  Here, $J_1$ is a Bessel
function, and $\gamma$ the Lorentz boost in the target frame.  For a
collider, $\gamma$ is related to the lab-frame boost $\gamma_{cm}$ by
$\gamma=2\gamma_{cm}^2 -1$. For heavy nuclei, the nuclear charge form
factor $F$ can be analytically modelled by the convolution of a hard
sphere with a Yukawa potential of range $0.7$ fm\cite{usrates}.

The probability of having a mutual Coulomb excitation in a collision with 
impact parameter $b$ is calculated along the same lines as in~\cite{BCW}. 
For low energy photons ($k < \sim$30~MeV in the target frame), the dominant 
photonuclear reaction is excitation of the target to a Giant Dipole 
Resonance (GDR). This collective excitation usually decays by single neutron 
emission, with multiple neutron emission also possible\cite{dissoc}.
Higher energy photons can excite
the nucleus to higher collective modes or excite individual nucleons (i.e.
via $\gamma p\rightarrow \Delta $).  More energetic interactions
generally lead to multiple neutron emission, sometimes accompanied by 
$\pi$ emission.

This paper will treat two cases: a general Coulomb excitation leading to the 
emission of any number of neutrons (Xn) and an excitation to a GDR followed by 
the emission of exactly one neutron (1n). The lowest-order probability for an 
excitation to any state which emits one or more neutrons ($Xn$) is 
\begin{equation}
P^1_{C(Xn)} (b) = \int dk {d^3n(b,k) \over dkd^2b}
\sigma_{\gamma A\rightarrow A^*}(k).
\label{eq:nx}
\end{equation}
Here, the superscript $1$ shows that this is the lowest order
probability.  The photoexcitation cross section $\sigma_{\gamma
A\rightarrow A^*}(k)$ is is determined by measurements at a wide range
of energies\cite{baltz}.

At small impact parameters, $P^1_{C}(b)$ can exceed $1$ and
so cannot be interpreted as a probability. Instead, it corresponds to 
the mean number of excitations. The excitation probability may be determined
by a unitarization procedure. The probability of having exactly N excitations 
follows a Poisson distribution
\begin{equation}
   P_N (b) = \frac{ \left( P^1 (b) \right)^N \, \exp{(- P^1 (b))} }{N!} \; .
\label{eq:pn}
\end{equation}
The probability of having at least one Coulomb excitation is then 
$P_{C(Xn)}(b)=1-\exp{(-P^1_{C(Xn)} (b))}$.

The probability of excitation followed by single neutron emission may
be similarly determined.  Here, the lowest order probability is
determined as in Eq. (\ref{eq:nx}), except that the photon energy
integration is truncated at the maximum GDR energy, and data on
photoemission of single neutrons is used, avoiding uncertainties about
the GDR branching ratios. For single neutron emission, there must be
a single $1n$ excitation, unaccompanied by higher excitations. So,
$P_{C(1n)}(b)=P^1_{C(1n)}(b) \exp{(-P^1_{C}(b))}$.

In mutual dissociation, each individual breakup occurs 
independently\cite{mutualdissoc}. The probability is thus the square of
the individual breakup probabilities, i.e. $P_{C(XnXn)}(b)=(P_{C(Xn)}(b))^2$
and $P_{C(1n1n)}(b)=(P_{C(1n)}(b))^2$.
Mutual dissociation is an experimental signature for low-$b$
events\cite{parkcity}.  When a nucleus breaks up, the neutron momentum
is largely unchanged, and so the neutrons are detectable in zero
degree calorimeters downstream of the interaction point\cite{BCW,chiu,zdc}.

Vector mesons are produced in peripheral heavy-ion interactions at 
ultra-relativistic energies through coherent interactions between 
the electromagnetic and nuclear fields\cite{usrates,parkcity}. A photon 
from the field of one nucleus fluctuates to a quark-antiquark pair and 
scatters elastically from the other nucleus, emerging as a vector meson. 
The cross section is sensitive to the vector meson-nucleon interaction 
cross section. The photon energy $k$ is related to the final state 
meson rapidity, $y=1/2\ln{(2k/M_V)}$, where $M_V$ is the vector 
meson mass. Since the photon energy spectrum depends on $b$, the meson 
rapidity distribution $d\sigma/dy$ also varies with $b$. A vector meson
can be produced at either ion and thus the two ions act as a two-source 
interferometer\cite{usinterf}. 

The probability for vector meson production $P_V(b)$ is similar to that
for nuclear excitation, except that $P_V(b)$ is small, so there is
no need for a unitarization process:
\begin{equation}
P_V (b) = \int dk {d^3n(b,k) \over dkd^2b} \sigma_{\gamma
A\rightarrow VA}(k).
\end{equation}
$P_V (b)$ is the probability for producing a vector meson at one nucleus. The 
total probability is obtained by doubling $P_V$ to take into account that
either nucleus can emit a photon. This factor of 2 was not present for
dissociation because Eq.~\ref{eq:pn} and our subseqent expressions for
$P_{C(Xn)}(b)$ and $P_{C(1n)}(b)$ were for excitation of a
specific nucleus.  The natural width of the $\rho^0$ is taken into account
in the integration over $k$~\cite{usrates}. 

Assuming that the sub-reactions are independent, the cross section to 
produce a vector meson accompanied by mutual dissociation is 
\begin{equation}
\sigma(AA\rightarrow A^*A^*V)\!=\!2\!\int\!d^2b\ P_V(b)
P_{XnXn}(b)\exp{(-P_H(b))}.
\label{eq:sigma}
\end{equation}
$P_{H}(b)$ is the mean number of projectile nucleons that interact at 
least once:
\begin{equation}
P_H (b) = \int  d^2\vec{r} \, T_A (\vec{r} - \vec{b}) 
\left( 1 -  \exp( - \sigma_{NN} T_B(\vec{r}) ) \right).
\end{equation}
The nuclear thickness function, $T(\vec{r})$ is calculated from the 
nuclear density distribution and using for the total nucleon-nucleon cross 
section $\sigma_{NN}$ = 52 mb (88 mb) at a center of mass energy of 200 GeV
(5.5 TeV) per nucleon pair\cite{BCW,usrates}.
The factor $\exp{(-P_H(b))}$ in Eq.~\ref{eq:sigma} thus ensures that the 
reaction is unaccompanied by hadronic interactions. 
For a solid-sphere nucleus model, the hadronic interaction probability 
is $1$ for $b<2R_A$ and is zero otherwise. Here, we calculate the 
interaction probability from a Glauber model. 

Table I gives the production cross sections and median impact
parameters $b_m$ for the different tags, as calculated 
from Eq.~\ref{eq:sigma}.  The $XnXn$ and $1n1n$ cross sections are about 
1/10 and 1/100 of the untagged cross sections, respectively. 

\begin{figure}
\setlength{\epsfxsize=0.8\textwidth}
\center{\epsffile{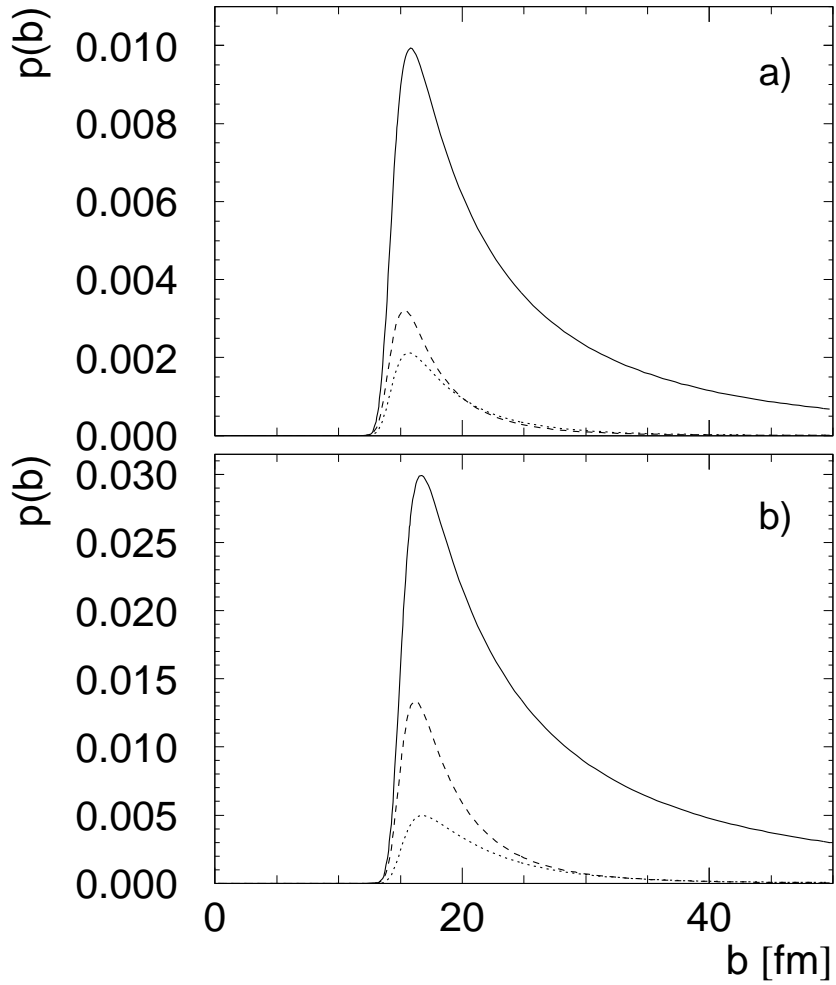}}
\caption[]{The probability of $\rho^0$ production with (a) gold beams
at RHIC and (b) lead beams at the LHC as a function of $b$, with
$XnXn$ (dashed curve) and $1n1n$ (dotted curve) and without nuclear
excitation (solid curve).  The $1n1n$ curve is multiplied by 10 to fit on
the plot.}
\label{probability}
\end{figure}

Fig. \ref{probability} compares the probability of $\rho$ production
as a function of impact parameter for $XnXn$ and $1n1n$ excitation and
also without requiring nuclear excitation for (a) gold-gold collisions
at a center of mass energy $\sqrt{s_{NN}} = 200$ GeV per nucleon, as
are found at the Relativistic Heavy Ion Collider (RHIC) at Brookhaven
National Laboratory, and (b) lead-lead collisions at $\sqrt{s_{NN}}
= 5.5$ TeV per nucleon as are planned at the Large Hadron Collider (LHC)
at CERN.  These curves were obtained by evaluating the integrand of
Eq. (\ref{eq:sigma}) at different $b$.  The $b$ distributions are very
different for tagged and untagged $\rho$ production; this is reflected 
in the vastly different $b_m$ in Tables 1 and 2. The $1n1n$ and $XnXn$
spectra are closer, except that $XnXn$ is more strongly peaked for
$b<20 fm$, likely reflecting the increased phase space for high-energy
excitations there.  This difference is reflected by a $\approx 10\%$
difference in $b_m$. With nuclear breakup, $b_m$ is almost
independent of the final state vector meson.

\begin{figure}
\setlength{\epsfxsize=0.8\textwidth}
\center{\epsffile{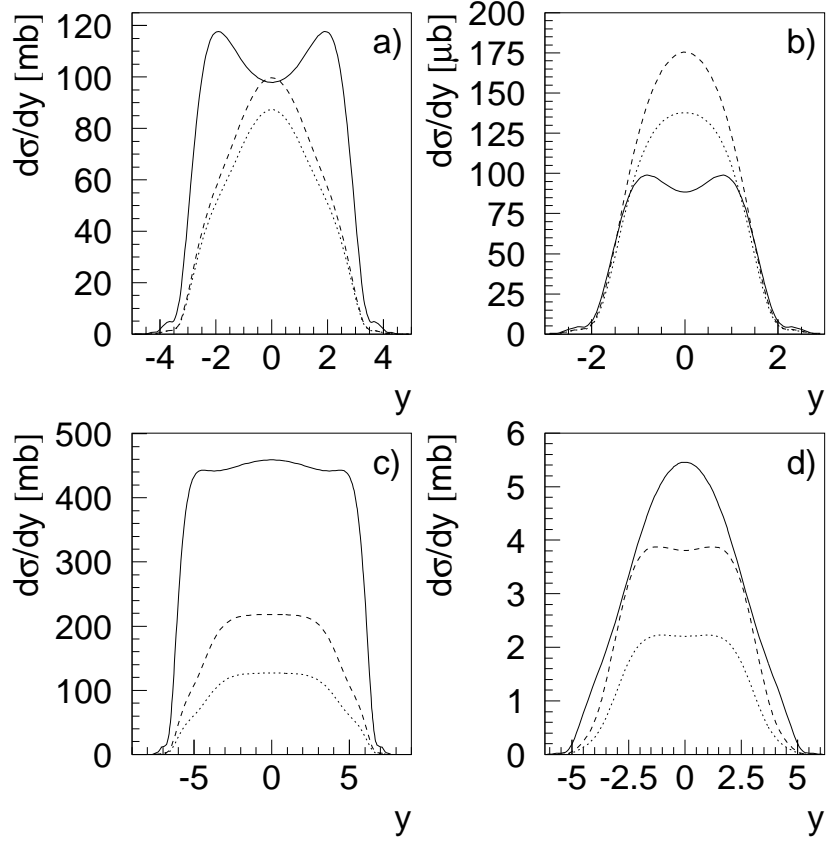}}
\caption[]{Rapidity spectrum $d\sigma/dy$ for (a) $\rho$ production at
RHIC, (b) $J/\psi$ production at RHIC, (c) $\rho$ production at LHC
and (d) $J/\psi$ production at the LHC.  The solid line is the total
production, the dashed line for $XnXn$, multiplied by 10, and the
dotted line is $1n1n$, multiplied by 100.}
\label{dndy}
\end{figure}

Figure \ref{dndy} shows the rapidity distribution $d\sigma/dy$ for
$\rho$ and $J/\psi$ production at RHIC and the LHC.  Spectra for
$XnXn$ and $1n1n$ breakup are shown, along with the untagged $d\sigma/dy$.
The $d\sigma/dy$ are symmetric
around $y=0$ because either nucleus can emit the photon. Since the
photon spectrum falls as $1/k$ and $y=1/2\ln{(2k/M_V)}$, the vector meson 
is usually in the hemisphere that the photon came from, with
large $|y|$ corresponding to low photon energies. The tagged spectra have 
a smaller $|y|$ than the untagged distribution, with a small difference 
between the $XnXn$ and $1n1n$ calculations.
Breakup preferentially selects collisions with smaller
impact parameters, with higher median photon energies, and so
the vector mesons are produced closer to mid-rapidity. 

The form factor for a nucleus in an excited (i.e. GDR) state may be 
different from ground state nuclei. This could conceivably affect the 
vector meson production, particularly the meson $p_t$ spectra. However, 
this should at most be a small effect and we neglect it here.

The meson $p_T$ spectrum also depends on $b$. For production at a
single source (target nucleus), the meson $p_T$ is the sum of the
photon and scattering $p_T$, which is largely independent of $b$. The
photon $p_T$ comes from the equivalent photon approximation, (2),
while the $p_T$ from the coherent scattering depends directly on the form
factor, Eq. (2) of Ref. \cite{usinterf}. The overall meson $p_T$ spectrum is
affected by interference from the two production sources (ions). The
two amplitudes add with a $b-$dependent phase factor
$\exp{(i\vec{p}_T\cdot\vec{b})}$. At mid-rapidity the two amplitudes
have the same magnitude but opposite sign, because of the negative 
parity of the vector meson, and 
\begin{equation}
\sigma(p_T,b) = \sigma_1(p_T,b) [1-cos{(\vec{p_T}\cdot\vec{b})}]
\label{eq:int}
\end{equation}
where $\sigma_1(p_T,b)$ is the cross section for emission from a
single source.  Of course, $\vec{b}$ is unknown, so Eq. (\ref{eq:int})
must be integrated over $\vec{b}$.  This integration washes out the
interference term in Eq. (\ref{eq:int}) except for $p_T <
\hbar/\langle b\rangle$, where the cross section is reduced.  As
$\langle b\rangle$ decreases, the interference extends to higher and
higher $p_T$. In this way, tagging affects the $p_T$ spectrum.

\begin{figure}
\setlength{\epsfxsize=0.8\textwidth}
\center{\epsffile{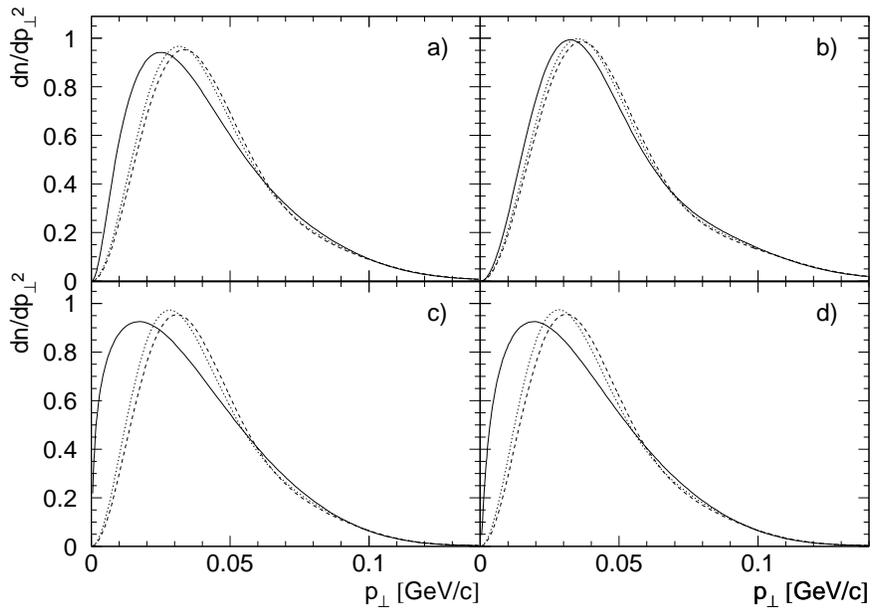}}
\caption[]{The transverse momentum spectrum $d^2\sigma/dp_t^2$ 
at mid-rapidity, $y=0$, for 
(a) $\rho$ production at RHIC, (b) $J/\psi$ production at RHIC, (c) $\rho$
production at LHC and (d) $J/\psi$ production at the LHC.  The solid
line is the total production, the dashed line for $XnXn$ and the
dotted line $1n1n$.  All of the curves assume that there is
interference and are normalized
so that
 without interference $dn/dp_T^2=1$ at $p_T=0$.}
\label{ptspectrum}
\end{figure}

Figure \ref{ptspectrum} compares the $p_T$ distributions,
$d^2N/dp_T^2$ for $\rho$ and $J/\psi$ production at RHIC and the LHC.
The solid line shows the untagged spectra, while the dashed and
dotted lines are for $XnXn$ and $1n1n$, respectively.  The
differences are moderate at RHIC and large at the LHC, where the
exclusive vector mesons can have much larger $b_m$ than the tagged sample.  
The different distributions can be
used to study interference under different conditions.

Although we have focused on $b-$tagging vector meson production, this
technique should also be useful for studying two-photon interactions 
at heavy ion colliders.  Here, $e^+e^-$ production is of special interest.  
At $b$ smaller than the electron Compton wavelength, 
$\lambda_C =$~386 fm, the fields are very strong, and multiple pairs 
production is enhanced over single pair production\cite{reviews}.  Mutual 
excitation could be used to select events with $b < \lambda_C$, and look for
enhanced multiple pair production. 

In conclusion, we have calculated the total cross sections and
rapidity and transverse momentum distributions for vector mesons
accompanied by nuclear breakup.  The presence of nuclear breakup is an
effective tag for events with smaller average $b$.  These events will
have different rapidity and $p_t$ distributions from the un-tagged
events, and can be used to explore the effects of different photon
spectra and $b$ distributions. 

This work was supported by the U.S. Department of Energy under 
Contracts No. DE-AC-03076SF00098 and DE-AC02-98CH10886, and  
by the Swedish Research Council (VR).

\begin{table}
\caption{Cross sections and median impact parameters
$b_m$, for production of vector
mesons.}
\begin{tabular}{lrrrrrr}
Meson 	&  \multicolumn{2}{c}{overall}  & \multicolumn{2}{c}{XnXn} & \multicolumn{2}{c}{1n1n} \\
 	& $\sigma$ [mb]  & $b_m$ [fm] & $\sigma$ [mb]  &  $b_m$ [fm] & $\sigma$ [mb]  &  $b_m$ [fm] \\ 
\hline
\multicolumn{7}{c}{Gold beams at RHIC ($\gamma_{cm} =$~108)}  \\
$\rho^0$& 590		& 46	 & 39	&  18 	& 3.5	& 19	\\
$\omega$& 59		& 46	 & 3.9	&  18	& 0.34	& 19	\\
$\phi$	& 39		& 38	 & 3.1	&  18	& 0.27	& 19	\\
J/$\psi$& 0.29		& 23	 & 0.044 & 17	& 0.0036& 18	\\
\hline
\multicolumn{7}{c}{Lead beams at LHC ($\gamma_{cm} =$~2940)}   \\
$\rho^0$& 5200		& 280	 & 210	& 19	& 12	& 22	\\
$\omega$& 490		& 290	 & 19	& 19	& 1.1	& 22	\\
$\phi$	& 460		& 220	 & 20	& 19	& 1.1	& 22	\\
J/$\psi$& 32		& 68	 & 2.5 	& 19	& 0.14	& 21	\\
\end{tabular}
\label{sigmaLHC}
\end{table}

\vfill\eject

\end{document}